\documentclass[11pt]{article}

\usepackage{amsmath,amsfonts,amssymb,bbm,epsfig}
\usepackage[textwidth=14.7cm,top=30mm,bottom=35mm]{geometry}
\usepackage{url}

\begin{document}

%
\def\be{\begin{equation}}
\def\ee{\end{equation}}
\def\bea{\begin{eqnarray}}
\def\eea{\end{eqnarray}}
 \newcommand{\ba}{\begin{eqnarray}}
 \newcommand{\ea}{\end{eqnarray}}
\def\rarr{\rightarrow}
\def\nn{\nonumber}
\renewcommand\slash[1]{\not \! #1}
\def\del{\partial}
\newcommand\nin{\noindent}

\begin{titlepage}
\begin{flushright}
\end{flushright}
\vspace{0.6cm}
\begin{center}
{\LARGE{\bf Central exclusive production at the LHC:}}\\[.2cm]
{\LARGE{\bf Remarks by the organisers on an}}\\[.2cm]
{\LARGE{\bf EMMI workshop}}\\
\end{center}
\vspace{1.2cm}
\begin{center}
{\bf \Large
Carlo Ewerz\,$^{a,b,1}$, Otto Nachtmann\,$^{a,2}$, Rainer Schicker\,$^{c,3}$}
\end{center}
\vspace{.2cm}
\begin{center}
$^a$
{\sl
Institut f\"ur Theoretische Physik, Universit\"at Heidelberg,\\
Philosophenweg 16, D-69120 Heidelberg, Germany}
\\[.5cm]
$^b$
{\sl
ExtreMe Matter Institute EMMI, GSI Helmholtzzentrum f\"ur Schwerionenforschung,\\
Planckstra{\ss}e 1, D-64291 Darmstadt, Germany}
\\[.5cm]
$^c$
{\sl 
Physikalisches Institut, Universit\"at Heidelberg,\\
Im Neuenheimer Feld 226, D-69120 Heidelberg, Germany}
\end{center}
\vfill
\begin{abstract}
\noindent
On February 6, 2019, an EMMI workshop on ``Central exclusive production at the LHC'' was held at Heidelberg. 
Here we make some remarks on the topics presented in the talks and the discussions of this meeting. We hope 
that this will be useful for further studies of central exclusive reactions. 
\end{abstract}
\vspace{5em}
\hrule width 5.cm
\vspace*{.5em}
{\small \noindent
$^1$ email: C.Ewerz@thphys.uni-heidelberg.de\\
$^2$ email: O.Nachtmann@thphys.uni-heidelberg.de\\
$^3$ email: schicker@physi.uni-heidelberg.de
}
\end{titlepage}

In the following the organisers present some remarks on the topics 
and ideas discussed at the EMMI workshop ``Central exclusive production at the LHC''. 
In this way we want to document 
things we have learnt from the workshop. We hope that this may 
be useful for colleagues studying such processes. All misunderstandings
and errors (which we try, of course, to avoid) are to be blamed on us, not on the 
speakers at the workshop.

\begin{figure}[h!t]
\begin{center}
\includegraphics[width=0.48\textwidth]{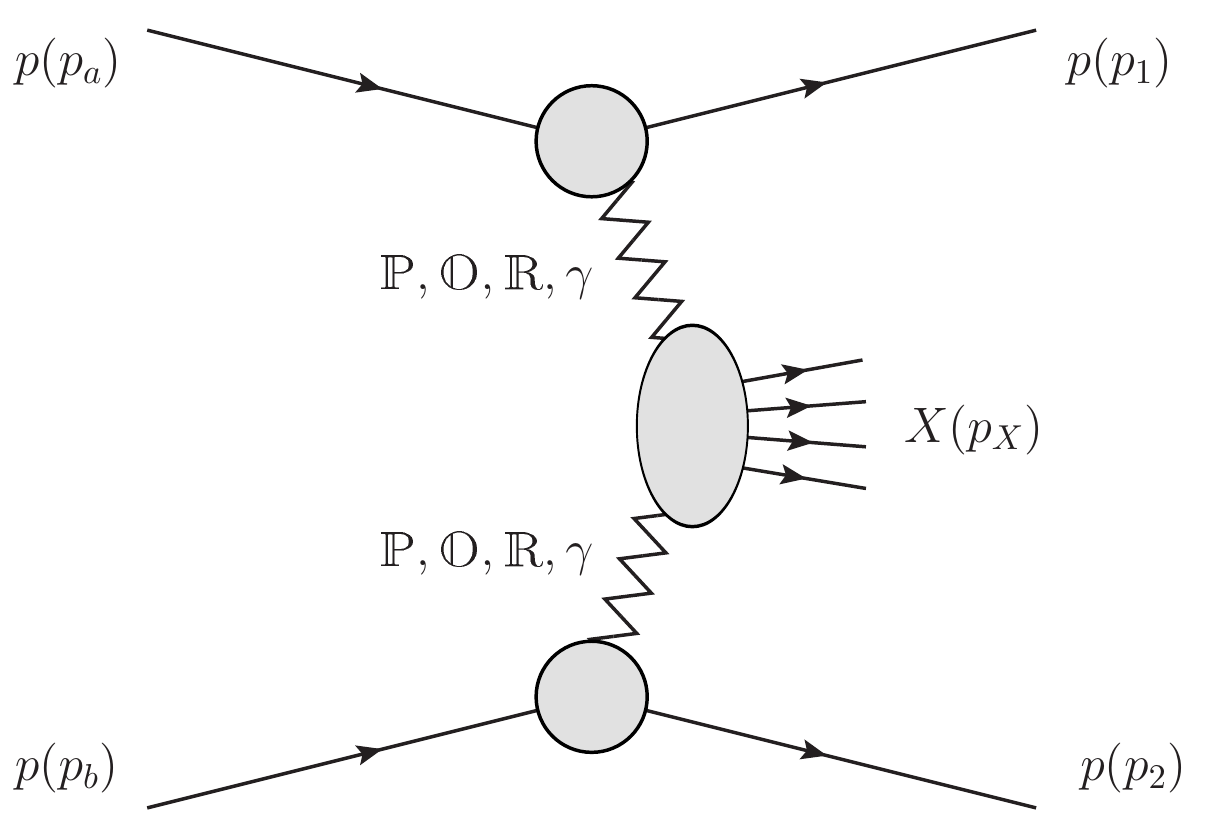}
\caption{Central exclusive production (CEP) in proton-proton collisions}
\label{Fig.1}
\end{center}
\end{figure}

The topic of the workshop was central exclusive production (CEP) at the
LHC, in particular the reaction 
\be\label{1}
p(p_a)+p(p_b)\longrightarrow p(p_1)+X(p_X)+p(p_2);
\ee
see Fig.\ \ref{Fig.1}. 
In place of the protons we can have nuclei $A$. We are always assuming
high centre-of-mass energies. Large rapidity gaps between $p(p_1)$ and $X$
as well as between $p(p_2)$ and $X$ are required. 
For a general review of CEP we refer to \cite{Albrow:2010yb}. 
We focussed on hadronic systems $X$ of low to moderate masses $m_X$. 
The exchange objects contributing in the reaction (\ref{1}) are the
pomeron ($\mathbbm{P}$), the odderon ($\mathbbm{O}$), 
reggeons ($\mathbbm{R} =f_{2R}, a_{2R},\omega_R,\rho_R)$ and the
photon ($\gamma$). 
The main questions to which the workshop was devoted were the following: 
\begin{itemize}
\item
How do the exchange objects couple to the external protons and
to the system $X$?
\item
What are good ways to look for odderon effects in CEP? 
\item
Which systems $X$ can be produced from the fusion of the respective exchange
objects and what are their properties?
\end{itemize}

The talks at the workshop tried to find answers to these questions. 
In the talks by Otto Nachtmann and Antoni Szczurek it was argued that
a successful effective description of the above exchanges has been
developed where the pomeron and the charge-conjugation $C=+1$
reggeons $f_{2R}$, $a_{2R}$ are treated as effective second-rank symmetric
tensor exchanges, in particular regarding their coupling to particles. 
The $C=-1$ exchanges ($\mathbbm{O}$, $\omega_R$, $\rho_R$)
are treated as effective vector exchanges. A central message of these
talks was that CEP offers excellent possibilities to look for effects of the
still elusive odderon. It was also shown that the determination of the
couplings of the system $X$ to the exchange objects would benefit
a lot if the final-state protons in (\ref{1}) could be measured. For further information
on the subject of these talks see 
\cite{Ewerz:2013kda,Lebiedowicz:2013ika,Bolz:2014mya,Lebiedowicz:2019jru,Britzger:2019lvc} 
and references therein. 

A different model for CEP was discussed by Laszlo Jenkovszky based on 
\cite{Fiore:2015lnz,Fiore:2017xnx,Jenkovszky:2018itd}. In this model
various resonances forming the system $X$ are treated together in a
Veneziano-type ansatz. This correlates the production rates of many resonances lying 
on a common regge trajectory. An interesting process which can be studied in CEP
is pomeron-pomeron fusion into glueballs. Certain glueball states are expected
to lie on the pomeron trajectory at positive values of its argument $t$.
Crucial for the identification of these states is the knowledge of the nonlinear
complex trajectory, interpolating between negative and positive values of $t$.
Its parameters are fitted to the scattering data, that is at negative $t$.
While the real part of the trajectory is almost linear, the recovery of the
imaginary part, determining the widths of the glueballs, is a highly non-trivial
problem. 

Anton Rebhan's talk dealt with central exclusive production of glueballs and
their decays. He stressed that even after extensive experimental searches
for glueballs there is, to date, no conclusive identification of any glueball
in the meson spectrum. Lattice gauge theory gives reliable predictions
only in pure Yang-Mills theory. In the real world, mixing of glueballs
with quark-antiquark states presents a great complication. An interesting
approach to calculate hadron properties relies on the AdS/CFT correspondence
or, more generally, holography. 
Anton Rebhan presented many results for the hadron spectrum using in the holographic 
framework the Witten-Sakai-Sugimoto model. For instance, detailed predictions
for the decay channels of a scalar glueball were given, which are consistent
with the interpretation of $f_0$(1710) as being predominantly glue. In this
approach glueballs up to and including spin 2 can be treated within the
supergravity approximation \cite{Brunner:2015yha,Brunner:2015oga,Brunner:2015oqa,Brunner:2018wbv}.
Higher spin states require a much more difficult string-theoretic treatment \cite{Imoto:2010ef}.

Maciej Trzebinski and Sergey Evdokimov gave talks on event generators for CEP, GenEx
and DRgen, respectively. Such event generators are essential for allowing detailed
comparisons between theory and experiment. We think that the development
of these generators is on a good way and very promising. At present single amplitudes for CEP 
reactions are treated. The next step will be to include all interference terms if
there are several amplitudes contributing. Finally, absorptive corrections will
have to be taken into account.
See \cite{Trzebinski:2017ikx,GenEx,Ataian:1991gn,AliRoot} and references therein. 

Alan Martin stressed in his talk that CEP reactions are very clean but that one pays the
price of huge reduction factors in the cross section due to absorption effects.
He identified eikonal and Sudakov suppression which, taken together, can reduce
cross sections calculated at the ``Born level'' by factors of order 10 and larger. 
Absorption also plays a big role and needs to be taken into account in the search
for odderon effects in high-energy proton-proton elastic scattering at small $|t|$. 
As a consequence, in his view, the existing evidence for the odderon from small $|t|$
proton-proton elastic scattering is not yet convincing. On the other hand, based 
on theoretical considerations in QCD,
the odderon should exist. Alan Martin discussed various CEP reactions which
may be suitable for odderon searches. However, experiments will have to deal with 
background problems which can sometimes be severe. He ended with the optimistic view that
``Experimentally the odderon is elusive but with experimental ingenuity and precision
it stands a good chance of being cornered.''
See \cite{Khoze:2017sdd,Khoze:2017swe,Khoze:2018kna,Harland-Lang:2018ytk,Harland-Lang:2018iur} 
and references therein. 

The workshop ended with a discussion session moderated by Peter Braun-Munzinger.

At \url{https://indico.gsi.de/event/8187} the pdfs of all talks presented at the workshop are available. 

Our conclusions from the talks and the discussions at the workshop are the following: 
\begin{itemize}
\item
CEP reactions will allow a detailed study of the couplings of the pomeron to various particles.
\item
CEP offers excellent possibilities to look for odderon-exchange effects. 
But it will require great experimental effort to realise this potential of CEP.
\item
CEP offers many opportunities to study the production of hadronic resonances. 
In particular, the search for and the characterisation of glueballs are prime topics.
\item
All these studies will benefit considerably if the complete final state of CEP can be measured
experimentally and if particle identification is available.
\end{itemize}

We give some examples: 
\begin{itemize}
 
\item 
The measurement of the transverse momenta of the outgoing protons
in (\ref{1}) will allow to determine the azimuthal angle $\phi_{pp}$ between
them. The angle $\phi_{pp}$ is important, e.g.\ for the determination of
the $\mathbbm{P} \mathbbm{P} X$ couplings.
 
\item
A measurement of the transverse momenta of the outgoing protons in (\ref{1})
is essential for a study of the so-called ``glueball-filter variable'' \cite{Close:1997pj}, 
given by the absolute value of the difference of these transverse momenta.
This variable is important for an identification of glueballs. Another important
characterisation of glueballs comes from the flavour composition of their 
decay channels which needs particle identification. 
 
\item
In $pp$, $Ap$, and $AA$ collisions there are good possibilities to study CEP
reactions with photon exchange. For this, good measurement possibilities at
small $p_T$, corresponding to small $|t|$, will be important. Only then will 
one be able to isolate and study photon exchange reactions. And, very importantly,
a search for odderon effects, where typically photon exchange is a competitor
which has to be separated, needs studies of $t$-dependences, including very small $|t|$.
 
\item
Even at LHC energies the reggeons $f_{2R}$, $a_{2R}$, $\omega_R$, $\rho_R$
will contribute to CEP. Kinematically one expects in general that their contribution
will be small for small pseudorapidities $\eta$ of the system $X$ in (\ref{1}).
But reggeon contributions will be non-negligible for large $|\eta|$ depending, 
of course, on the reaction at hand. Thus, a study of the couplings of reggeons
to the central systems $X$ in (\ref{1}) will need good measurement capabilities
for large $|\eta|$; $|\eta|$ up to 4 or 5 could be sufficient. 
 
\item
To conclude we want to say that the topics studied at the workshop were
mostly QCD topics. Perturbative QCD can be applied to some hard diffractive
central production processes, for instance to central jet production.
The emphasis of this workshop, however, was on reactions belonging to the
realm of non-perturbative QCD. At present the theory for these CEP reactions
uses effective models where results can only partly be checked against ``exact''
QCD results, for instance, from lattice gauge theory. We can envisage in the
future a very fruitful interplay of experiment and theory advancing our
knowledge of QCD in this interesting non-perturbative regime of very high energies
and small momentum transfers. 
 \end{itemize}
 
\nin
{\bf Acknowledgments:}
The authors would like to thank the ExtreMe Matter Institute EMMI at the
GSI Helmholtzzentrum f\"ur Schwerionenforschung, Darmstadt, Germany, for the support of this workshop. 
Furthermore, our thanks go to all speakers of the workshop for their excellent talks and to all 
participants, present at the workshop or connected via video, for their contributions to the discussions.

\end{document}